\documentclass[10pt]{article}\usepackage{utopia}
\usepackage{fullpage}
\usepackage[margin=1in]{geometry}
\usepackage{authblk, amsmath, amssymb, amsthm, setspace, bm, mathtools, multirow, breqn, bigints, enumitem, comment}
\usepackage{natbib}
\usepackage{hyperref}\hypersetup{colorlinks=true, linkcolor=blue, citecolor=blue, urlcolor=blue}
\usepackage{graphicx, float, subcaption}
\usepackage{algorithm}
\usepackage{caption}
\usepackage[title]{appendix}

\providecommand{\keywords}[1]{\small\textbf{Keywords:} #1}

\DeclareMathOperator{\diag}{diag}
\DeclareMathOperator{\ind}{\mathbb{I}}
\newtheorem{theorem}{Theorem}

\newtheorem{example}[theorem]{Example}

\title{\bf Robust Bayesian Model Averaging for Linear Regression Models With Heavy-Tailed Errors}

\author{Shamriddha De \footnote{PhD Candidate, Department of Statistics and Actuarial Science, The University of Iowa, Iowa City, USA}
\ and
Joyee Ghosh \footnote{Associate Professor, Department of Statistics and Actuarial Science, The University of Iowa, Iowa City, USA}}

\date{}

\begin{document}

\maketitle

\setstretch{1}
\begin{abstract} \noindent
Our goal is to develop a Bayesian model averaging technique in linear regression models that accommodates heavier tailed error densities than the normal distribution. Motivated by the use of the Huber loss function in the presence of outliers, the Bayesian Huberized lasso with hyperbolic errors has been proposed and recently implemented in the literature (\cite{Park:Case:2008, Kawa:Hash:2023}). Since the Huberized lasso cannot enforce regression coefficients to be exactly zero, we propose a fully Bayesian variable selection approach with spike and slab priors to address sparsity more effectively. The shapes of the hyperbolic and the Student-$t$ density functions are different. Furthermore, the tails of a hyperbolic distribution are less heavy compared to those of a Cauchy distribution. Thus, we propose a flexible regression model with an error distribution encompassing both the hyperbolic and the Student-$t$ family of distributions, along with an unknown tail heaviness parameter, that is estimated based on the data. It is known that the limiting form of both the hyperbolic and the Student-$t$ distributions is a normal distribution. 
We develop an efficient Gibbs sampler with Metropolis Hastings steps for posterior computation. Through simulation studies and analyses of real datasets, we show that our method is competitive with various state-of-the-art methods.
\\~\\
\keywords{Generalized hyperbolic distribution, Huberized Bayesian lasso, Hyperbolic distribution, Markov chain Monte Carlo model composition (MC$^3$), Spike and slab priors, Student-$t$ distribution.}
\end{abstract}

\singlespacing

\section{Introduction} \label{sec:intro}

Searching for suitable variable selection methods from a Bayesian perspective has been a venerable inquiry in the arena of linear regression models,
with \cite{Clyd:Geor:2004, Ohar:Sill:2009, Garc:Mart:2013, Fort:Garc:Stee:2018} providing a few reviews, among others. A large class of those techniques relies upon a presumption of normal errors. Such an assumption, however, compromises with the robustness of the obtained estimates towards extreme observations. This motivates the necessity of devising a method for Bayesian variable selection that enhances the normality assumption by flexibly adapting to an unknown degrees of tail heaviness.

Normal errors correspond to a quadratic loss function, which is sensitive to unusual observations. The Huber loss function (\cite{Hube:1973, Hube:Ronc:2011}) has been proposed as an alternative to the quadratic loss function. It is defined as
\begin{equation}
    \mathcal{L}_{\mathrm{Huber}}(a; c) = 
    \begin{cases}
        a^2/2, & \text{if } |a| \leq c, \\
        c(|a| - c/2), & \text{otherwise},
    \end{cases}
    \quad a \in \mathbb{R},
\label{eq:huberloss}
\end{equation}
where $c > 0$ is a tuning parameter. From the above definition, it can be seen that the Huber loss function is quadratic in a region around zero and linear outside that region. This makes the Huber loss function less sensitive to extreme observations compared to the quadratic loss function. However, Bayesian posterior computation under the Huber loss function is not straightforward. 

\cite{Park:Case:2008} suggested the hyperbolic loss function as an approximation to the Huber loss function. The hyperbolic loss function is given by
\begin{equation}
    \mathcal{L}_{\mathrm{Hyperbolic}}(a; \eta, \rho^2) = \sqrt{\eta (\eta + a^2/\rho^2)},
    \quad a \in \mathbb{R},
\end{equation}
where $\eta > 0$ and $\rho^2 > 0$ are parameters that are typically unknown. The hyperbolic loss function corresponds to assuming a hyperbolic distribution (\cite{Barn:1978}) for errors. The hyperbolic distribution can be expressed as a scale mixture of normal distributions (\cite{Gnei:1997}), which leads to more tractable Bayesian computation, compared to using the Huber loss function. \cite{Park:Case:2008} proposed the preliminary idea of the `Bayesian Huberized lasso' to robustify the traditional Bayesian lasso with normal errors, although without any explicit implementation. Furthermore, they assumed the tail heaviness parameter $\eta$ to be fixed in their discussion of the method. However, owing to its vital role as an unknown shape parameter in the error density, $\eta$ demands a prior. Recently, \cite{Kawa:Hash:2023} implemented the Bayesian Huberized lasso with a continuous prior on $\eta$. This leads to an intractable full conditional distribution for $\eta$, since $\eta$ appears as an argument in a modified Bessel function in the likelihood. \cite{Kawa:Hash:2023} developed an approximate Gibbs sampler by approximating the full conditional distribution of $\eta$.

The Bayesian Huberized lasso uses continuous Laplace priors for the regression coefficients, which assist in shrinkage but do not facilitate variable selection. In this article, we propose a mixture of a point mass at zero and a continuous prior on the regression coefficients to allow exclusion of variables from a model. Many authors have used this prior and its variants for variable selection in regression models with normal errors, such as \cite{Geor:Mccu:1993, Geor:McCu:1997, Clyd:Ghos:Litt:2011, Ghos:Clyd:2011, Hans:2010, Hans:2011, Ghos:Ghat:2015, Rock:Geor:2018, Nie:Rock:2023}, to name a few. If there are $p$ covariates, the  Bayesian Huberized lasso can be thought of as operating in a single model framework with all variables, that is, under the full model. In contrast, our framework has $2^p$ models corresponding to the potential inclusion/exclusion of each of the $p$ variables. This makes the posterior computation significantly more challenging in our setup, especially due to the tail heaviness parameter, to which we also assign a prior. We tackle this computational challenge by specifying a discrete prior with a flexible but finite support for the tail heaviness parameter $\eta$. This facilitates efficient posterior computation due to the existence of a tractable full conditional distribution for $\eta$.

The shapes of the hyperbolic and the Student-$t$ density functions are distinct. As a result, some heavy-tailed members of the hyperbolic family of distributions have the potential to generate extreme observations more frequently than some heavy-tailed members of the Student-$t$ family, such as the Cauchy distribution. However, the Cauchy distribution has heavier tails
than any hyperbolic density. This implies that, while the Cauchy distribution may produce extreme observations much less frequently, they could be much more extreme in magnitude than those produced under the hyperbolic distribution. Moreover, the limiting form of both the hyperbolic and the Student-$t$ distributions is a normal distribution. Therefore, we propose a model in which the errors follow either a hyperbolic or a Student-$t$ distribution. With this flexible modeling strategy, we can take advantage of the Huber loss function via the hyperbolic distribution component of the error model. However, if necessary, the model can also seamlessly adapt to the heavier tails of the Student-$t$ distribution or to the thinner tails of the normal distribution, as dictated by the data. A model with Student-$t$ errors has been developed by \cite{Gram:Pant:2010}, who also put spike and slab priors on the regression coefficients. They used a continuous prior for the tail heaviness parameter and developed a reversible jump MCMC sampler for posterior computation. In Sections \ref{sec:simul} and \ref{sec:data}, we compare our method to those of \cite{Gram:Pant:2010} and \cite{Kawa:Hash:2023}, and demonstrate that our method is competitive with these state-of-the-art methods.

While the aforementioned papers are most closely related to our work, there are many other interesting papers on Bayesian robust regression. We provide a brief review of a few of them here. Our proposed method as well as the previously reviewed literature rely on a traditional fully Bayesian posterior formulation based on a likelihood and a prior. \cite{Jian:Tann:2008} proposed an alternative approach using a Gibbs posterior based on a risk function, which bypasses the need to model the data directly. They illustrated that this alternative approach can be useful under model misspecification using examples for binary data. \cite{Kund:Duns:2014} proposed non-normal error distributions via semiparametric location mixture of normals, unlike scale mixture of normals, as in our case. They focused on $g$-priors and showed that their method enjoys model selection consistency. Motivated by applications in gene environment interaction, \cite{Ren:Zhou:Li:Ma:etal:2023} proposed a model with Laplace errors and spike and slab priors for Bayesian variable selection. Instead of changing the likelihood, \cite{Hans:Peru:Wang:2023} developed an alternative strategy by tuning a Zellner's $g$-prior to obtain optimal predictions in the presence of outliers. 

Before proceeding further, the organization of the present article is outlined as follows. In Section \ref{sec:hem}, we review the hyperbolic error model. In Section \ref{sec:htem}, we introduce our proposed mixture model using hyperbolic and Student-$t$ errors. In Section \ref{sec:simul}, we carry out a simulation study to compare the performance of the proposed mixture model with some competing models. We analyze the Boston housing dataset and NBA player salaries in Section \ref{sec:data}. Concluding remarks and possible directions for future work are stated in Section \ref{sec:conc}. Additionally, for clarity on parametrization, the probability density functions of some distributions that are used throughout the article are provided in an Appendix.

\section{Hyperbolic Error Model} \label{sec:hem}

In this section, we first briefly review the hyperbolic distribution. Next, we consider a regression model with hyperbolic errors and spike and slab priors, which serves as one of the components for our proposed mixture model described in Section \ref{sec:htem}.

\subsection{Review of the Hyperbolic Distribution} \label{sec:hem_motivation}

Mathematically, the hyperbolic density is defined as
\begin{equation}
    f_{\mathrm{Hyperbolic}}(x |\theta, \alpha, \delta, \mu, \phi) = \dfrac{\theta}{2 \alpha \delta K_1(\delta\theta)} e^{-\alpha\sqrt{\delta^2 + (x-\mu)^2} + \phi(x-\mu)},
    \quad -\infty < x < \infty,
\label{eq:hyperbdensityorig}
\end{equation}
where $K_1$ is a modified Bessel function with index 1, $\theta = \sqrt{\alpha^2 - \phi^2}$, $\mu \in \mathbb{R}, \alpha > 0, -\alpha < \phi < \alpha$ and $\delta > 0$. A reparametrized symmetric (about zero) version of the density is obtained by substituting $\phi = \mu = 0$, $\alpha = \theta = \sqrt{\dfrac{\eta}{\rho^2}}$ and $\delta = \sqrt{\eta\rho^2}$ in \eqref{eq:hyperbdensityorig}, thereby yielding
\begin{equation}
    f_{\mathrm{Hyperbolic}}(x | \eta, \rho^2) = \dfrac{1}{2 \sqrt{\eta \rho^2} K_1(\eta)} e^{-\left\{ \eta(\eta + x^2/\rho^2) \right\}^{1/2}},
    \quad -\infty < x < \infty,
\label{eq:hyperbdensity}
\end{equation}
with shape parameter $\eta > 0$ and scale parameter $\rho^2 > 0$. Let us denote the density as $\mathrm{Hyperbolic}(\eta, \rho^2)$. This density forms a special case of a larger class of distributions, viz., the generalized hyperbolic distribution (\cite{Barn:1978}), whose symmetric members can be represented as a normal scale mixture with a generalized inverse Gaussian mixing density. We highlight a special case of Example \ref{eg:hyperb_normalscalemix} by \cite{Gnei:1997}, that explicitly states this result.
\begin{example}[\cite{Gnei:1997}]
    Let $A$ and $a$ be random variables such that $A|a^2 \sim \mathrm{N}(0, a^2)$ and $a^2 \sim \mathrm{GIG}(1, \eta/\rho^2, \eta\rho^2)$. Then $A$ is distributed as $\mathrm{Hyperbolic}(\eta, \rho^2)$.
\label{eg:hyperb_normalscalemix}
\end{example}
\begin{figure}[!ht]
\centering
    \includegraphics[scale=0.75]{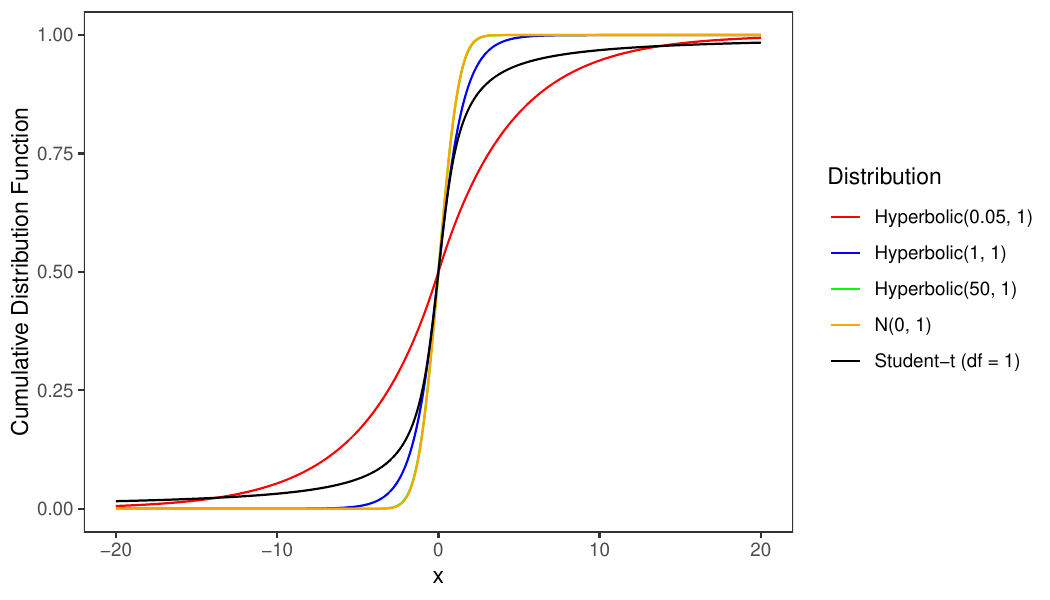}
\caption{Cumulative distribution functions of different symmetric densities.}
\label{fig:tailcomp_cdf}
\end{figure}
The hyperbolic density in \eqref{eq:hyperbdensity} has a heavier tail than the normal density, but is lighter-tailed than a Cauchy distribution. This is illustrated in Figure \ref{fig:tailcomp_cdf}, where the cumulative distribution functions of hyperbolic distributions (for different choices of $\eta$ and $\rho^2 = 1$) are compared with those for standard normal and Student $t$-distributions with one degree of freedom (that is, the standard Cauchy distribution). Furthermore, we note that the $\mathrm{Hyperbolic}(\eta=50, \rho^2=1)$ distribution is almost identical to the standard normal distribution, which provides useful information for specifying a prior for $\eta$ in (\ref{eq:htem_prior_etagrid}), in Section \ref{sec:htem_model}. Additionally, the hyperbolic distribution approaches a Laplace distribution as the shape or tail heaviness parameter $\eta$ tends to zero \citep{Barn:1977}.

\subsection{Model Formulation} \label{sec:hem_model}

Let us consider a linear regression model with $p$ covariates as
\begin{equation}
    \bm{Y} = X \bm{\beta} + \bm{\epsilon},
\label{eq:hem_regmod}
\end{equation}
where $n$ is the sample size, $\bm{Y} = (y_1, y_2, \dots, y_n)^\top$ denotes the $n$-dimensional response vector, $X$ is the $n \times p$ design matrix, $\bm{\beta} = (\beta_1, \beta_2, \dots, \beta_p)^\top$ corresponds to the $p$-dimensional vector of regression coefficients, and $\bm{\epsilon}$ is the $n$-dimensional vector of independent and identically distributed hyperbolic errors with a probability density function \eqref{eq:hyperbdensity}. The covariates as well as the response variables are centered about their respective means and scaled by their standard deviations, such that incorporating an intercept term in model \eqref{eq:hem_regmod} is not necessary. Besides, we also incorporate variable selection uncertainty. In particular, we define a $p$-dimensional binary vector $\bm{\gamma} = (\gamma_1, \gamma_2, \dots, \gamma_p)^\top$, where $\gamma_j$ ($j = 1, 2, \dots, p$) takes the value one or zero depending on whether the $j$th covariate is included in or excluded from the model. For example, the vectors $\bm{\gamma} = (1, 1, \dots, 1)^\top$ and $\bm{\gamma} = (0, 0, \dots, 0)^\top$ represent the full model with all covariates and the null model with no covariates, respectively. When $\gamma_j = 0$, we assume the corresponding regression coefficient, $\beta_j = 0$, for $j = 1, 2, \dots, p$. Thus, under a variable selection framework, some of the components of $\bm{\beta}$ can be exactly zero. Let $\bm{\Gamma} = \{0, 1\}^p$ denote the model space consisting of all the $2^p$ possible models. For every model $\bm{\gamma}$ in the model space $\bm{\Gamma}$, we define $\bm{\beta}_{\bm{\gamma}}$ to be the $p_{\bm{\gamma}}$-dimensional vector of the non-zero components in $\bm{\beta}$, and let $X_{\bm{\gamma}}$ signify the matrix of corresponding columns of the design matrix $X$. Using the normal scale mixture representation of the hyperbolic density stated in Example \ref{eg:hyperb_normalscalemix}, we can rewrite model \eqref{eq:hem_regmod} in the following form, which assists in simplifying posterior computation. 
\begin{align}
    \bm{Y} | \bm{\gamma}, \bm{\beta}, \sigma^2_1, \sigma^2_2, \dots, \sigma^2_n
    & \sim
    \mathrm{N} (X_{\bm{\gamma}} \bm{\beta}_{\bm{\gamma}}, \Sigma),
    \label{eq:hem_lik}
\\
    \sigma^2_i | \rho^2, \eta
    & \stackrel{\mathrm{ind}}{\sim}
    \mathrm{GIG}(1, \eta/\rho^2, \eta\rho^2),
    \quad i = 1, 2, \dots, n.
    \label{eq:hem_prior_sigma2}
\end{align}
For the unknown parameters, we specify the following priors:
\begin{align}
    \beta_j | \gamma_j, \rho^2, \tau^2
    & \stackrel{\mathrm{ind}}{\sim}
    \mathrm{N}(0, \rho^2 \tau^2)\ind{(\gamma_j = 1)} + \delta_{\{0\}}(\beta_j)\ind{(\gamma_j = 0)},
    \quad j = 1, 2, \dots, p,
    \label{eq:hem_prior_beta}
\\
    \tau^2
    & \sim
    \mathrm{IGamma}(\nu/2, \nu/2),
    \label{eq:hem_prior_tau2}
\\
    \rho^2
    & \sim
    \mathrm{IGamma}(a, b),
    \label{eq:hem_prior_rho2}
\\
    \eta
    & \sim
    \mathrm{DiscUniform}(\mathcal{S}_\eta),
    \label{eq:hem_prior_eta}
\\
    \gamma_j | \tilde{\pi}
    &\stackrel{\mathrm{ind}}{\sim}
    \mathrm{Bernoulli}(\tilde{\pi}),
    \quad j = 1, 2, \dots, p,
    \label{eq:hem_prior_gamma}
\\
    \tilde{\pi}
    & \sim
    \mathrm{Beta}(s_1, s_2),
    \label{eq:hem_prior_pi}
\end{align}
where $\nu > 0$, $a > 0$, $b > 0$, $s_1 > 0$, $s_2 > 0$, $\Sigma = \diag{(\sigma^2_1, \sigma^2_2, \dots, \sigma^2_n)}$, $\mathcal{S}_\eta$ is a finite set of positive real numbers, $\delta_{\{0\}}(\cdot)$ represents a point mass at zero and $\ind{(\cdot)}$ denotes the indicator function. It is easy to observe that marginalizing over $\tau^2$ from \eqref{eq:hem_prior_beta} using \eqref{eq:hem_prior_tau2} leads to a multivariate-$t$ prior with $\nu$ degrees of freedom on the non-zero regression coefficients, while retaining point mass priors on the zero coefficients. In this manner, we put spike and slab priors on the regression coefficients. The aforementioned priors and likelihood lead to closed form full conditional distributions, from which samples can be drawn with relative ease. Thus, we develop a Gibbs sampler for posterior computation.

We close our deliberation on the hyperbolic error model through a summary of its significance and pitfalls. To deal with heavy-tailed data, a plausible contender is the Student-$t$ error model. However, the simulation study in Section \ref{sec:simul_hem_htem_tem} reveals the usefulness of the hyperbolic error model, when the true data generating model is hyperbolic. Our simulation study suggests that the hyperbolic model can have superior performance in estimation of regression coefficients in such a case. These results provide another motivation for using the hyperbolic error model, in addition to its relationship with the Huber loss. The convergence of the hyperbolic density to a Laplace density for small values of $\eta$ (\cite{Barn:1977}) provides yet another motivation for including the former distribution. On the other hand, the Student-$t$ distribution can achieve a higher degree of tail heaviness compared to the hyperbolic distribution. Besides, the two distributions have contrasting tail structures. While very large observations can be sampled from the hyperbolic density more frequently than the Student-$t$ density, the Student-$t$ distribution has the potential to generate values that are even more extreme than the hyperbolic distribution. At the same time, a behavior similar to a normal distribution can be achieved by both the distributions for large values of the tail heaviness parameter. Accordingly, both the hyperbolic and the Student-$t$ error models have some benefits and drawbacks, entailing a probe for a more befitting error model.

\section{Mixture of Hyperbolic and Student-$t$ Error Models} \label{sec:htem}

It has been discussed in Section \ref{sec:hem} that despite having heavier tails than a normal distribution, the hyperbolic distribution has thinner tails than some members of the family of Student-$t$ distributions. However, there are situations where a hyperbolic error model provides better estimation than a Student-$t$ error model, like the one shown in Section \ref{sec:simul_hem_htem_tem}. The two distributions also differ in shapes, as mentioned in Section \ref{sec:hem}. This suggests that it would be appealing to have a model that accounts for the characteristics of both distributions. Inspired by this idea, we propose a linear regression model, where the errors follow a mixture of hyperbolic and Student-$t$ distributions. We incorporate variable selection uncertainty in our model, via spike and slab priors on the regression coefficients.

\subsection{Model Formulation} \label{sec:htem_model}

We return to the linear regression model \eqref{eq:hem_regmod} with $p$ covariates and $n$ observations introduced in Section \ref{sec:hem_model}. While the notational connotations remain unchanged, the error distribution becomes different. As mentioned at the beginning of this section, we consider a mixture of hyperbolic and Student-$t$ errors as
\begin{equation}
    \epsilon_i | \alpha, \eta, \rho^2
    \stackrel{\mathrm{iid}}{\sim}
    \mathrm{Hyperbolic}(\eta, \rho^2) \ind{(\alpha = 0)}
    + \mathrm{t}(\eta, \rho^2) \ind{(\alpha = 1)},
    \quad
    i = 1, 2, \dots, n,
\label{eq:htem_errdist}
\end{equation}
where $\mathrm{t}(\eta, \rho^2)$ refers to a Student-$t$ distribution with location parameter zero, degrees of freedom parameter $\eta$, and scale parameter $\rho^2$, and the parameter $\alpha \in \{0, 1\}$ denotes an indicator function for choosing the error distribution between hyperbolic and Student-$t$. Equation \eqref{eq:htem_errdist} implies that conditional on $\alpha = 0$, all the errors are generated from the hyperbolic density, that is 
$\epsilon_i | \alpha=0, \eta, \rho^2 \stackrel{\mathrm{iid}}{\sim} \mathrm{Hyperbolic}(\eta, \rho^2)$, $i = 1, 2, \dots, n$, 
while conditional on  $\alpha = 1$, all the errors are generated from the Student-$t$ density, that is 
$\epsilon_i | \alpha=1, \eta, \rho^2 \stackrel{\mathrm{iid}}{\sim} \mathrm{t}(\eta, \rho^2)$, $i = 1, 2, \dots, n$.

Using the normal scale mixture representations of the hyperbolic and Student-$t$ distributions, we can express the above regression model in a more computationally convenient form as
\begin{align}
    \bm{Y} | \bm{\gamma}, \bm{\beta}, \sigma^2_1, \sigma^2_2, \dots, \sigma^2_n
    & \sim
    \mathrm{N} (X_{\bm{\gamma}} \bm{\beta}_{\bm{\gamma}}, \Sigma) ,
    \label{eq:htem_lik}
\\
    \sigma^2_i | \alpha, \eta, \rho^2
    & \stackrel{\mathrm{iid}}{\sim}
    \mathrm{GIG}(1, \eta/\rho^2, \eta\rho^2) \ind{(\alpha = 0)}
    + \mathrm{IGamma}(\eta/2, \eta\rho^2/2) \ind{(\alpha = 1)},
    \quad i = 1, 2, \dots, n ,
    \label{eq:htem_prior_sigma2}
\end{align}
where $\Sigma = \diag{(\sigma^2_1, \sigma^2_2, \dots, \sigma^2_n)}$. Under such a model, we are tasked with a simultaneous learning of the error density, tail heaviness and variable selection. To account for the tail heaviness, we fall back on the strategy of putting a discrete uniform prior on $\eta$, as used in Section \ref{sec:hem}, while a conditional Bernoulli prior is assigned to the error indicator $\alpha$. The remaining parameters retain their previous priors stated in Section \ref{sec:hem}. In other words, we introduce the priors
\begin{align}
    \alpha | \omega
    & \sim
    \mathrm{Bernoulli}(\omega);
    \label{eq:htem_prior_alpha}
\\
    \omega
    & \sim
    \mathrm{Beta}(m_1, m_2);
    \label{eq:htem_prior_omega}
\end{align}
where $m_1 > 0$ and $m_2 > 0$, and maintain the prior structure in \eqref{eq:hem_prior_beta} through \eqref{eq:hem_prior_pi} for the rest of the unknown parameters.

The challenge in the prior specification now shifts to the selection of a suitable grid $\mathcal{S}_{\eta}$ for the prior of $\eta$ in \eqref{eq:hem_prior_eta}. It is important to note that the parameter $\eta$ has a two-fold implication. When $\alpha = 0$, $\eta$ serves as the shape parameter under hyperbolic errors. On the other hand, $\eta$ indicates the degrees of freedom for Student-$t$ errors for $\alpha = 1$. While $\eta$ controls the tail heaviness under both hyperbolic and Student-$t$ error models, its interpretation changes when the error density changes. Therefore, it is crucial for the prior to encompass a wide range of tail heaviness under both error distributions. Figure \ref{fig:tailcomp_cdf} shows a drastic reduction in tail heaviness of the hyperbolic density from $\eta = 0.05$ to $\eta = 1$, followed by a slow and gradual subsidence thereafter. When the value of $\eta$ becomes large enough and approaches 50, the cumulative density functions of the hyperbolic and the standard normal densities are nearly indistinguishable. Based on the above information, we specify a support for $\eta$, that consists of a dense grid of values in the interval $(0, 1]$, and a relatively sparse grid when $\eta > 1$. Furthermore, it is well known that a Student-$t$ distribution with more than 30 degrees of freedom behaves similar to a normal distribution, while low degrees of freedom (about 3 or less) correspond to significantly heavy tails. Accordingly, we choose
\begin{equation}
    \mathcal{S}_{\eta} = \{0.05, 0.1, 0.2, 0.3, \dots, 0.9, 1, 2, 5, 10, 20, 50\},
\label{eq:htem_prior_etagrid}
\end{equation}
as the prior support of $\eta$, which provides an extensive coverage of varying degrees of tail heaviness for both the error distributions.

\subsection{Posterior Computation} \label{sec:htem_posterior}

Equipped with the model and priors in Section \ref{sec:htem_model}, we can derive all the full conditional distributions in closed form. Therefore, we can develop a Gibbs sampling algorithm to simulate from the joint posterior distribution 
$p(\bm{\gamma}, \bm{\beta}, \sigma^2_1, \sigma^2_2, \dots, \sigma^2_n, \tau^2, \rho^2, \alpha, \eta, \tilde{\pi}, \omega | \bm{Y})$. It is worth mentioning that a naive implementation of the Gibbs sampler can lead to poor mixing, so we perform several block updates to improve mixing. One possibility is to draw the pairs $(\gamma_j, \beta_j)$ component-wise, for $j = 1, \dots, p$, conditional on  ${\gamma_1, \beta_1, \dots, \gamma_{j-1}, \beta_{j-1}, \gamma_{j+1}, \beta_{j+1}, \dots, \gamma_p, \beta_p}$, other parameters, and the observed data.
Since this updating scheme can lead to poor mixing, we carry out a block update for $\bm{\beta}$ and $\bm{\gamma}$. Conditional on other parameters, we integrate out $\bm{\beta}$, and draw $\bm{\gamma}$ using a Metropolis-Hastings step, with an add-delete proposal, as in the Markov chain Monte Carlo model composition (MC$^3$) algorithm (\cite{Madi:York:1995, Hoet:Madi:Raft:Voli:1999}). Given a model $\bm{\gamma}$, the MC$^3$ algorithm selects one component of $\bm{\gamma}$ at random, say $\gamma_k$, and the proposed model $\bm{\gamma}^*$ is formed by flipping $\gamma_k$ to $1-\gamma_k$. 
The proposed model $\bm{\gamma}^*$ is then accepted with the appropriate Metropolis-Hastings acceptance probability. It may be noted that this proposal
randomly adds a variable to the current model or drops one. We also perform block updates for $\alpha$, $\eta$ and $(\sigma^2_1, \sigma^2_2, \dots, \sigma^2_n)^\top$, in that order. A block update for $\eta$ is possible due to the carefully chosen prior with finite support, and is a critical step for improved mixing. The full conditional distributions are given as follows.
\begin{align}
    \begin{split}
        p(\bm{\gamma} | \bm{Y}, \sigma^2_1, \sigma^2_2, \dots, \sigma^2_n, \tau^2, \rho^2, \tilde{\pi})
        & \propto
        \left[
            \dfrac{|A_{\bm{\gamma}}|^{-1/2}}{(\tau^2 \rho^2)^{p_{\bm{\gamma}}/2}}
            \exp{\left( \dfrac{1}{2} \bm{Y}^\top \Sigma^{-1} X_{\bm{\gamma}} A_{\bm{\gamma}}^{-1} X_{\bm{\gamma}}^\top \Sigma^{-1} \bm{Y} \right)}
            \ind(\bm{\gamma} \neq \bm{0})
            + \ind(\bm{\gamma} = \bm{0})
        \right] \\
        & \quad \times
          \tilde{\pi}^{p_{\bm{\gamma}}} (1 - \tilde{\pi})^{p - p_{\bm{\gamma}}},       
    \end{split}
    \label{eq:htem_post_gamma}  
\\
    \bm{\beta} | \bm{Y}, \bm{\gamma}, \sigma^2_1, \sigma^2_2, \dots, \sigma^2_n, \tau^2, \rho^2
    & \sim
    \mathrm{N} \left(
        A_{\bm{\gamma}}^{-1} X_{\bm{\gamma}}^\top \Sigma^{-1} \bm{Y},
        A_{\bm{\gamma}}^{-1}
    \right),
    \label{eq:htem_post_beta}
\end{align}
\begin{align}
    \begin{split}
        \rho^2 | \bm{\gamma}, \bm{\beta}, \sigma^2_1, \sigma^2_2, \dots, \sigma^2_n, \tau^2, \alpha, \eta
        & \sim
        \mathrm{GIG} \left(
            -\left(a + n + \dfrac{p_{\bm{\gamma}}}{2}\right),
            \eta \sum_{i=1}^n \dfrac{1}{\sigma^2_i},
            2b + \dfrac{\bm{\beta}_{\bm{\gamma}}^\top \bm{\beta}_{\bm{\gamma}}}{\tau^2} + \eta \sum_{i=1}^n \sigma^2_i
        \right)
        \ind{(\alpha = 0)} \\
        & \quad
        + \mathrm{GIG} \left(
            \dfrac{n\eta - p_{\bm{\gamma}} - 2a}{2},
            \eta \sum_{i=1}^n \dfrac{1}{\sigma^2_i},
            2b + \dfrac{\bm{\beta}_{\bm{\gamma}}^\top \bm{\beta}_{\bm{\gamma}}}{\tau^2}
        \right)
        \ind{(\alpha = 1)},
    \end{split}
    \label{eq:htem_post_rho2}
\end{align}
\begin{align}
    \begin{split}
        p(\alpha | \bm{Y}, \bm{\gamma}, \bm{\beta}, \rho^2, \omega)
        & \propto
        \left[ (1-\omega) \sum_{\eta \in \mathcal{S}_\eta}
            \prod_{i=1}^n \dfrac{1}{2 \sqrt{\eta \rho^2} K_1(\eta)} \exp{\left(-\sqrt{\eta \left(\eta + \dfrac{\epsilon_i^2}{\rho^2}\right)}\right)}
        \right] \ind{(\alpha = 0)} \\
        & \quad
        + \left[ \omega \sum_{\eta \in \mathcal{S}_\eta}
            \prod_{i=1}^n \dfrac{\Gamma((\eta+1)/2)}{\Gamma(\eta/2)} \dfrac{\eta^{\eta/2}}{\sqrt{\pi\rho^2}} \left(\eta + \dfrac{\epsilon_i^2}{\rho^2}\right)^{-(\eta+1)/2}
        \right] \ind{(\alpha = 1)},
    \end{split}    
    \label{eq:htem_post_alpha}
\\
   \begin{split}
        p(\eta | \bm{Y}, \bm{\gamma}, \bm{\beta}, \rho^2, \alpha)
        & \propto
        \left[
            \left\{
                \prod_{i=1}^n \dfrac{1}{2 \sqrt{\eta \rho^2} K_1(\eta)} \exp{\left(-\sqrt{\eta \left(\eta + \dfrac{\epsilon_i^2}{\rho^2}\right)}\right)}
            \right\}
        \ind{(\eta \in \mathcal{S}_\eta)} \right] \ind{(\alpha = 0)} \\
        & \quad
        + \left[
            \left\{
                \prod_{i=1}^n \dfrac{\Gamma((\eta+1)/2)}{\Gamma(\eta/2)} \dfrac{\eta^{\eta/2}}{\sqrt{\pi\rho^2}} \left(\eta + \dfrac{\epsilon_i^2}{\rho^2}\right)^{-(\eta+1)/2}
            \right\}
        \ind{(\eta \in \mathcal{S}_\eta)} \right] \ind{(\alpha = 1)},
    \end{split}
    \label{eq:htem_post_eta}
\\
    \begin{split}
        \sigma^2_i | \bm{Y}, \bm{\gamma}, \bm{\beta}, \rho^2, \alpha, \eta
        & \stackrel{\mathrm{ind}}{\sim}
        \mathrm{GIG} \left(
            \dfrac{1}{2},
            \dfrac{\eta}{\rho^2}, \epsilon_i^2 + \eta\rho^2
        \right)
        \ind{(\alpha = 0)}
        + \mathrm{IGamma} \left(
            \dfrac{\eta+1}{2},
            \dfrac{\epsilon_i^2 + \eta\rho^2}{2}
        \right)
        \ind{(\alpha = 1)}, \\
        & \omit \hfill \text{$i = 1, 2, \dots, n$},
    \end{split}
    \label{eq:htem_post_sigma2}
\end{align}
\begin{align}
    \tau^2 | \bm{\gamma}, \bm{\beta}, \rho^2
    & \sim
    \mathrm{IGamma} \left(
        \dfrac{\nu + p_{\bm{\gamma}}}{2},
        \dfrac{\bm{\beta}_{\bm{\gamma}}^\top \bm{\beta}_{\bm{\gamma}}}{2\rho^2} + \dfrac{\nu}{2}
    \right),
    \label{eq:htem_post_tau2}
\\
    \tilde{\pi} | \bm{\gamma}
    & \sim
    \mathrm{Beta} \left(
        s_1 + p_{\bm{\gamma}},
        s_2 + p - p_{\bm{\gamma}}
    \right),
    \label{eq:htem_post_pi}
\\
    \omega | \alpha
    & \sim
    \mathrm{Beta} \left(
        m_1 + \alpha,
        m_2 + 1 - \alpha
    \right),
    \label{eq:htem_post_omega}
\end{align}
where $A_{\bm{\gamma}} = X_{\bm{\gamma}}^\top \Sigma^{-1} X_{\bm{\gamma}} + \dfrac{1}{\tau^2 \rho^2} I_{p_{\bm{\gamma}}}$, $p_{\bm{\gamma}} = \sum_{j=1}^p \bm{\gamma}_j$, and ${\epsilon}_i = y_i - \bm{x}^{(\bm{\gamma}) \top}_i \bm{\beta}_{\bm{\gamma}}$, for $i = 1, 2, \dots, n$. With a current model state at $\bm{\gamma}$, the acceptance probability of a proposed model $\bm{\gamma}^*$ in the Metropolis-Hastings step is computed as
\begin{equation}
    \pi_a = \min{\left\{
        1,
        \dfrac{p(\bm{\gamma}^* | \bm{Y}, \sigma^2_1, \sigma^2_2, \dots, \sigma^2_n, \tau^2, \rho^2, \tilde{\pi})}{p(\bm{\gamma} | \bm{Y}, \sigma^2_1, \sigma^2_2, \dots, \sigma^2_n, \tau^2, \rho^2, \tilde{\pi})}
    \right\}}.
    \label{eq:htem_post_modap}
\end{equation}
The sampling scheme is traced out in Algorithm \ref{algo:htem_sampler}.
\begin{algorithm}
\caption{Gibbs sampler with Metropolis-Hastings step for the HTEM}
\label{algo:htem_sampler}
    Given a current state, the next state is generated as follows:
    \begin{enumerate}
        \item Draw $\tau^2$ from the full conditional distribution in \eqref{eq:htem_post_tau2}.
        \item Draw $\rho^2$ from the full conditional distribution in \eqref{eq:htem_post_rho2}.
        \item
        \begin{enumerate}[label = (\alph*)]
            \item Draw $\alpha$ from the conditional distribution in \eqref{eq:htem_post_alpha}.
            \item Draw $\eta$ from the conditional distribution in \eqref{eq:htem_post_eta}.
            \item Draw $(\sigma^2_1, \sigma^2_2, \dots, \sigma^2_n)^\top$ from the full conditional distribution in \eqref{eq:htem_post_sigma2}.
        \end{enumerate}
        \item Draw $\tilde{\pi}$ from the full conditional distribution in \eqref{eq:htem_post_pi}.
        \item Draw $\omega$ from the full conditional distribution in \eqref{eq:htem_post_omega}.
        \item
        \begin{enumerate}[label = (\alph*)]
            \item Update $\bm{\gamma}$ using a Metropolis-Hastings step with the acceptance probability $\pi_a$ defined in \eqref{eq:htem_post_modap}.
            \item Draw $\bm{\beta}$ from the full conditional distribution in \eqref{eq:htem_post_beta}.
        \end{enumerate}
    \end{enumerate}
\end{algorithm}

The niceties of the algorithm enable a seamless movement between the two error models while allowing regression coefficients to be exactly zero, within a simple Gibbs sampling framework. Moreover, large values of $\eta$ for either of the mixed error densities tend to mimic a relatively light tailed structure like a normal error model. At this juncture, we make a salient observation. If the value of $\alpha$ is fixed at 0, the mixture model becomes identical to the hyperbolic error model in Section \ref{sec:hem}, and the algorithm will generate samples from the posterior of the hyperbolic model only. On the other hand, by enforcing $\alpha = 1$, we can sample from the posterior of the Student-$t$ error model. As such, the proposed mixture model potentially emerges as a flexible layout, with hyperbolic and Student-$t$ error models as special cases.

\section{Simulation Study} \label{sec:simul}

In this section, we compare the performance of our proposed mixture model (HTEM) in Section \ref{sec:htem} with some other candidate methods through several simulation studies. For this purpose, we consider a regression model as
\begin{equation}
    y_i = \beta_0 + \beta_1 x_{i1} + \beta_2 x_{i2} + \dots + \beta_p x_{ip} + \epsilon_i,
    \quad
    i = 1, 2, \dots, n,
\end{equation}
where $n = 100$ and $p = 100$. The covariates $x_{ij}$s ($i = 1, 2, \dots, n$; $j = 1, 2, \dots, p$) are generated independently from a $\mathrm{Uniform}(-2, 2)$ distribution. A total of six scenarios have been studied by varying the true values of regression coefficients and the true error distribution. In particular, the following scenarios are considered, under a sparse framework, where only a few coefficients among $\beta_1, \beta_2, \dots, \beta_{p}$ are non-zero:
\begin{itemize}
    \item Strong signals: $\beta_0 = 2$, $\beta_1 = \beta_2 = \beta_5 = \beta_7 = \beta_{10} = 3$, and the rest are set to zero.
    \item Mixed (strong and weak) signals: $\beta_0 = 2$, $\beta_1 = 0.5$, $\beta_2 = 1.5$, $\beta_3 = 2$, $\beta_4 = -3$, and the rest are set to zero.
\end{itemize}
The true error distributions used for the simulations are $\mathrm{Hyperbolic}(\eta=0.5, \rho^2=2)$, $\mathrm{N}(0, 2)$ (normal with mean 0 and variance 2), and $\mathrm{t}(\eta=2.1, \rho^2=1)$.
Combining everything together, the six studied scenarios are listed below:
\begin{enumerate}[label = (\Roman*)]
    \item \label{simul:mod_sh} Strong signals with errors drawn independently from $\mathrm{Hyperbolic}(0.5, 2)$,
    \item \label{simul:mod_sn} Strong signals with errors drawn independently from $\mathrm{N}(0, 2)$,
    \item \label{simul:mod_st} Strong signals with errors drawn independently from  $\mathrm{t}(2.1, 1)$, 
    \item \label{simul:mod_wh} Mixed signals with errors drawn independently from $\mathrm{Hyperbolic}(0.5, 2)$,
    \item \label{simul:mod_wn} Mixed signals with errors drawn independently from $\mathrm{N}(0, 2)$,
    \item \label{simul:mod_wt} Mixed signals with errors drawn independently from $\mathrm{t}(2.1, 1)$.
\end{enumerate}
We carry out the comparison with respect to estimation and predictive performances in two parts. At first, we compare our proposed method the HTEM with its special cases: the hyperbolic error model (HEM) and the Student-$t$ error model (TEM). Thereafter, we compare the performance of HTEM with some state-of-the-art methods for Bayesian variable selection and/or Bayesian robust regression.

\subsection{Comparison of the HTEM With its Special Cases, the HEM and the TEM} \label{sec:simul_hem_htem_tem}

As discussed in Section \ref{sec:htem}, the HTEM reduces to the hyperbolic error model (HEM) studied in Section \ref{sec:hem} and the Student-$t$ error model (TEM), by fixing the value of the error indicator $\alpha$ at 0 and 1 respectively. In this section, we illustrate the flexibility of the HTEM over the HEM or the TEM alone, through simulation studies under Scenarios (I), (III), (IV) and (VI), in which the true distribution of the errors is hyperbolic or Student-$t$.

We first state our choice of hyperparameters used to implement the three methods.  We take $\nu = 1$ degree of freedom for the multivariate-$t$ prior on $\bm{\beta}_{\bm{\gamma}}$, the vector of non-zero regression coefficients, to have a reasonably heavy-tailed prior. For the inverse gamma prior on the scale parameter $\rho^2$, the hyperparameters are taken as $a = 2.1$ and $b = 0.1$, to have most of the prior mass concentrated below 1, as the response variables are standardized to have scale 1. The prior on the inclusion probability $\tilde{\pi}$ is taken to be $\mathrm{Beta}(s_1 = 1, s_2 = \sqrt{p})$, in order to favor a sparse model with a relatively small number of nonzero regression coefficients, as recommended by \cite{Li:2013}. We use a $\mathrm{Beta}(m_1 = 1, m_2 = 1)$  prior on $\omega$, to give equal weight to the HEM and the TEM, apriori. The prior for the tail heaviness parameter $\eta$ has been specified in \eqref{eq:htem_prior_etagrid}, in the earlier section.

All the Gibbs samplers are executed for 100,000 iterations, after discarding the first 10,000 iterations as burn in. The regression coefficients are estimated using the posterior medians of the MCMC samples due to an inherent robustness of the median over the mean. Using these estimates, the root-mean-squared errors (RMSEs) of the regression coefficients are calculated as $\sqrt{\frac{1}{p+1} \sum_{j=0}^p (\beta_j - \hat{\beta}_j)^2}$, with $\hat{\beta}_j$ being the estimated posterior median of $\beta_j$. We also compute the relative RMSEs, defined as dividing the RMSE for a method by the lowest RMSE among all the studied methods. The relative RMSE will be close to 1, for a method that is frequently the best, in terms of RMSE. The RMSEs of signals (non-zero regression coefficients) and noise variables (regression coefficients with a true value of zero) are also examined separately.

\begin{figure}[!ht]
\centering
    \includegraphics[scale=0.75]{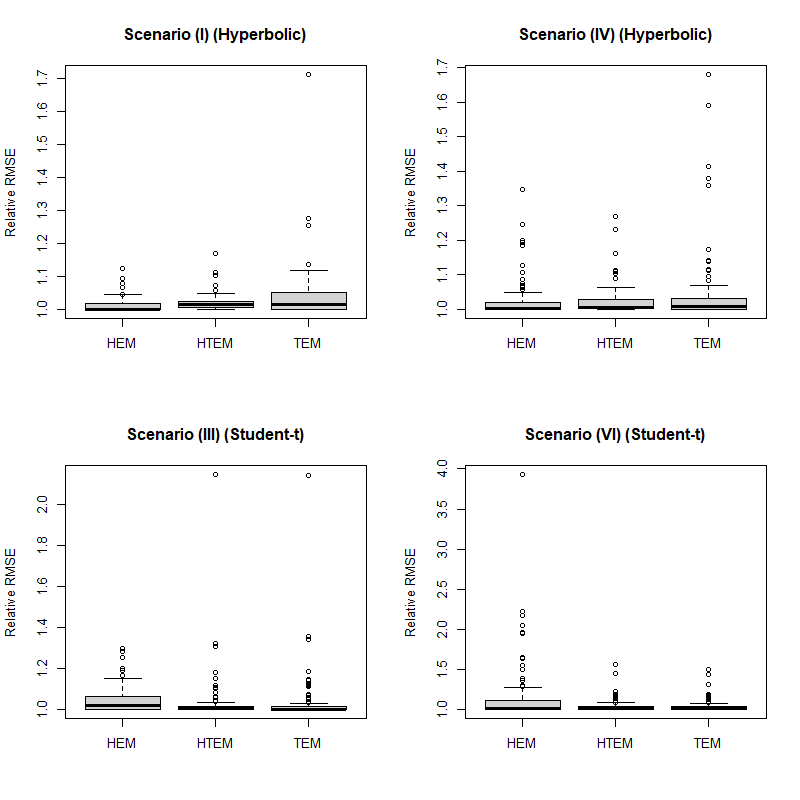}
\caption{Relative RMSEs of all the regression coefficients for the HTEM and its special cases under different scenarios. Relative RMSE of any method is its RMSE relative to that of the method having the smallest RMSE. The boxplots are based on 100 replicates.}
\label{fig:rmse_hem_htem_tem}
\end{figure}

Based on 100 replicates, the boxplots of overall relative RMSEs are presented in Figure \ref{fig:rmse_hem_htem_tem}.
A general trend can be observed from these boxplots. The HEM usually produces the smallest RMSEs under true hyperbolic errors, while the TEM seems to outperform the rest under true Student-$t$ errors. The RMSEs for the HTEM appear to be in between those for the HEM and the TEM, being slightly closer to the true error model. The HTEM is required to estimate the true error density, unlike the true error model. In other words, the HTEM is assigned a more difficult task than the true error model, which explains why it may be outperformed by the true error model. The fact that the HTEM tends to have relative RMSEs closer to the true error model is promising, and underscores the importance of having a mixture model, instead of fitting only one of them.

\begin{figure}[!ht]
\centering
    \includegraphics[scale=0.75]{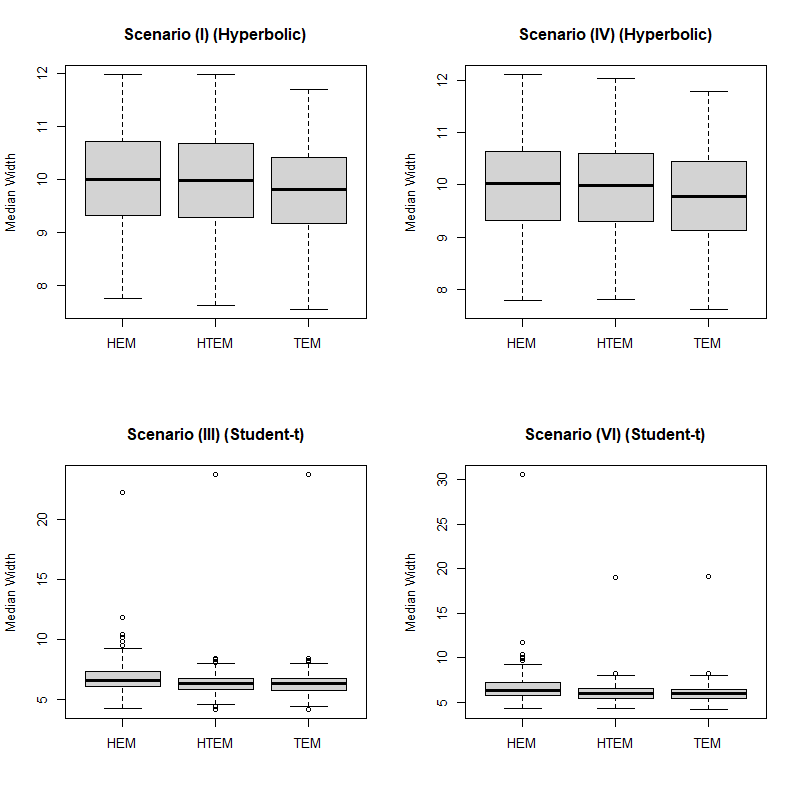}
\caption{Median widths of prediction intervals for the HTEM and its special cases under different scenarios. The boxplots are based on 100 replicates.}
\label{fig:pimw_hem_htem_tem}
\end{figure}

We also compare the predictive performance of the HTEM, the HEM and the TEM. For each of the four scenarios, we generate a new test dataset of 1,000 data points. Using the previously obtained 100,000 MCMC samples based on the training dataset, we construct 90\% prediction intervals with the quantiles of the posterior predictive distribution. The empirical coverage probabilities and the boxplots of median widths of the prediction intervals are laid out in Table \ref{tab:picp_hem_htem_tem} and Figure \ref{fig:pimw_hem_htem_tem} respectively.
\begin{table}
\centering
\caption{Empirical coverage probabilities of 90\% prediction intervals for the HTEM and its special cases under different scenarios, averaged over 100 replicates.}
\begin{tabular}{cccc}
    \hline
    & HEM & HTEM & TEM \\
    \hline
    Scenario (I) (Hyperbolic) & 0.899 & 0.897 & 0.892 \\
    Scenario (III) (Student-t) & 0.908 & 0.898 & 0.897 \\
    Scenario (IV) (Hyperbolic) & 0.896 & 0.895 & 0.888 \\
    Scenario (IV) (Student-t) & 0.902 & 0.891 & 0.890 \\
    \hline
\label{tab:picp_hem_htem_tem}
\end{tabular}
\end{table}
The observations in regard to the predictive performance are nearly consistent across all the simulation scenarios. Overall, each of the three methods maintain a coverage close to 90\%. On a more rigid scrutiny, the HEM maintains the largest empirical coverage with the largest median width, while the TEM provides the smallest of such values. Similar to the RMSEs of regression coefficients, the HTEM typically lies in between the HEM and the TEM, with an inclination towards the true model.

At this juncture, we observe that the HTEM strikes a balance between the HEM and the TEM. The performance of the HTEM tends to be closer to the true model. For real data, when the true error distribution is unknown, using the HTEM seems a natural solution, instead of choosing one among the HEM and the TEM. Therefore,  we focus on the more flexible HTEM solely for the rest of our study, and compare it with existing methods for Bayesian variable selection.

\subsection{Comparison of the HTEM With Existing Methods} \label{sec:simul_htem_others}

This section ventures to explore the advantages of the HTEM vis-\`a-vis some prevailing candidate methods. The Huberized lasso (\cite{Park:Case:2008}), being a stemming entity for our proposed method, asks for consideration in our comparison. The Bayesian Huberized lasso (HBL) of \cite{Kawa:Hash:2023} uses an approximate Gibbs sampler. \cite{Kawa:Hash:2023} put a gamma prior on the shape parameter $\eta$. This gives rise to a full conditional distribution for $\eta$, that is not available in closed form. Thus, \cite{Kawa:Hash:2023} approximate the full conditional of $\eta$ by a gamma distribution. However, in our comparison, we refrain from using such approximation and replace the same with our own discrete uniform prior on the grid in \eqref{eq:htem_prior_etagrid}. We also use the inverse gamma prior in \eqref{eq:hem_prior_rho2} on the scale parameter $\rho^2$ instead of an improper prior proposed by \cite{Kawa:Hash:2023}. Such an approach is deployed to ensure a parity among the compared methods. The HBL uses Laplace priors on the regression coefficients corresponding to the lasso. With our implementation of the HBL, the comparison of the HBL with the HTEM boils down to comparing 1) Laplace versus spike and slab priors on regression coefficients, and 2) hyperbolic errors in the HBL, versus a mixture of hyperbolic and Student-$t$ errors in the HTEM. In addition to the HBL, we consider the normal and the Student-$t$ error models proposed by \cite{Gram:Pant:2010}, which perform variable selection using spike and slab priors, through reversible jump MCMC algorithms. Both the techniques by \cite{Gram:Pant:2010} are implemented using the \texttt{monomvn} package (\cite{Gram:Rmonomvn}) in R, and we denote them by MONON and MONOT for the normal and Student-$t$ models respectively.

\begin{figure}[!ht]
\centering
    \includegraphics[scale=0.75]{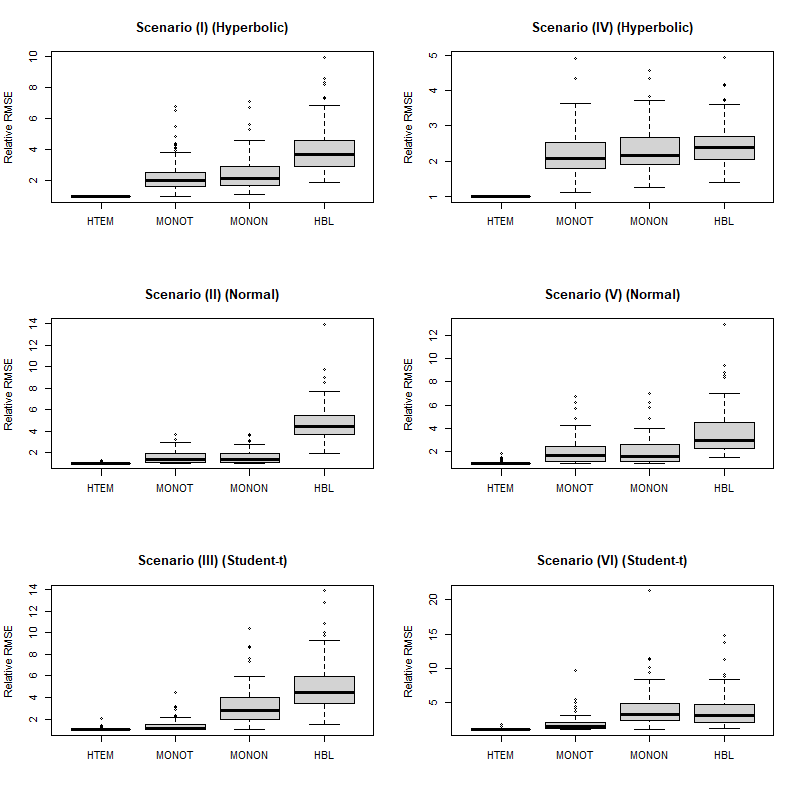}
\caption{Relative RMSEs of all the regression coefficients for the HTEM and the other studied competing methods under different scenarios. Relative RMSE of any method is its RMSE relative to that of the method having the smallest RMSE. The boxplots are based on 100 replicates.}
\label{fig:rmse_htem_others}
\end{figure}

\begin{figure}[!ht]
\centering
    \includegraphics[scale=0.75]{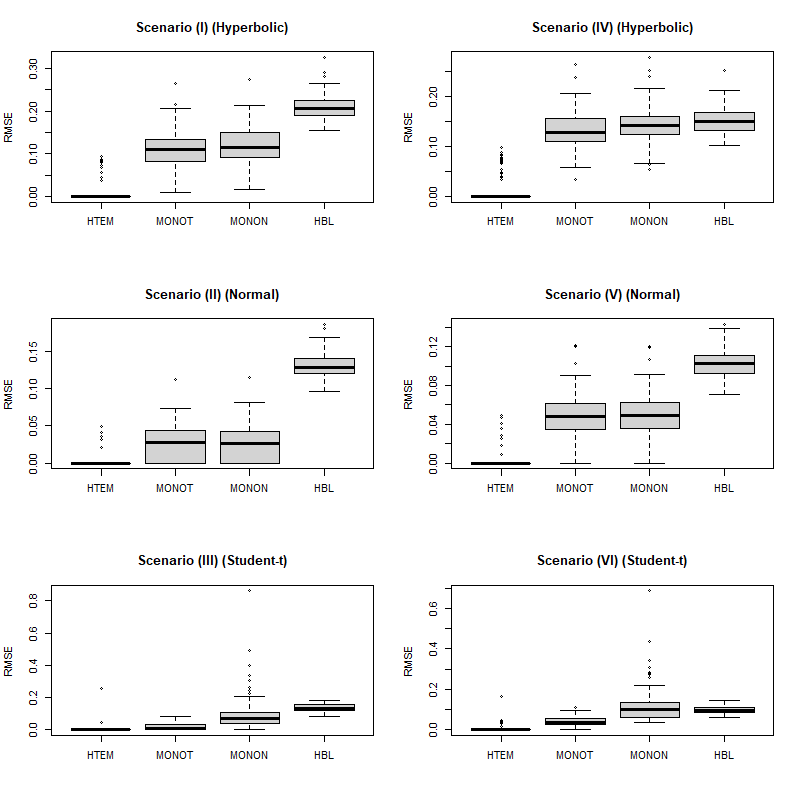}
\caption{RMSEs of noise variables (regression coefficients with true value zero) for the HTEM and the other studied competing methods under different scenarios. The boxplots are based on 100 replicates.}
\label{fig:rmse_zero_htem_others}
\end{figure}

\begin{figure}[!ht]
\centering
    \includegraphics[scale=0.75]{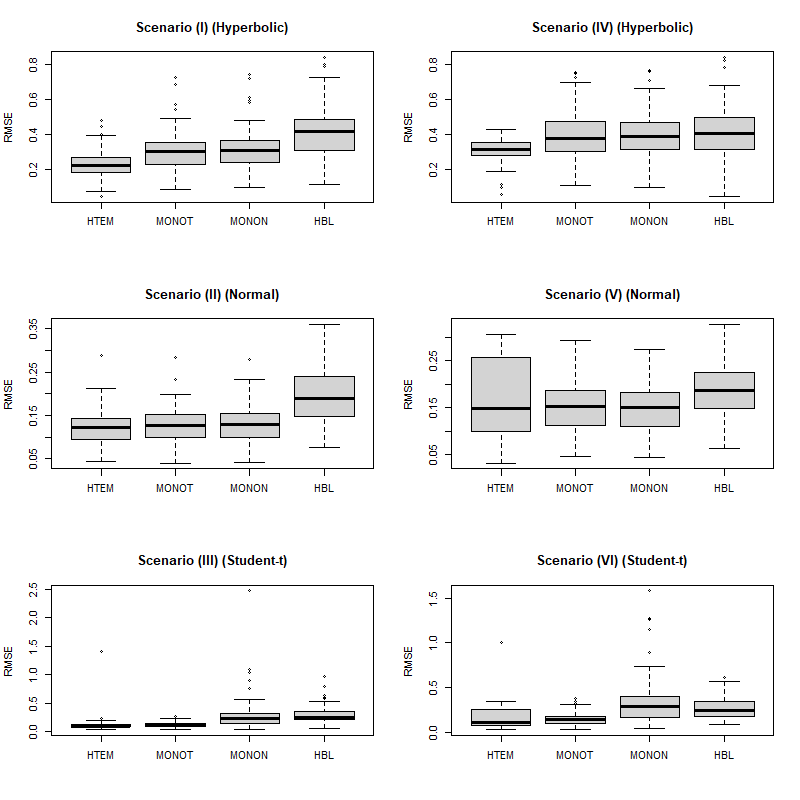}
\caption{RMSEs of signals (non-zero regression coefficients) for the HTEM and the other studied competing methods under different scenarios. The boxplots are based on 100 replicates.}
\label{fig:rmse_nonzero_htem_others}
\end{figure}

To carry out the present comparison, we look at all the six scenarios mentioned earlier, pertaining to the three error densities and two types of signals. We use the same hyperparameter values in Section \ref{sec:simul_hem_htem_tem}, and run the Gibbs samplers for 100,000 iterations, after a burn in of 10,000 samples, as before. For comparison, we employ the same metrics as in Section \ref{sec:simul_hem_htem_tem}. The overall RMSEs and relative RMSEs are summarized in Table \ref{tab:rmse_htem_others} and Figure \ref{fig:rmse_htem_others}. Across all the scenarios, the HTEM shows substantial gain in overall RMSE. In view of the sparseness involved in our simulation setups, the HBL typically yields large RMSEs with Laplace priors. Under heavy-tailed errors, the MONON also results in large RMSEs. As the closest robust competitor with spike and slab priors, the MONOT usually provides smaller RMSEs compared to the MONON and the HBL, but significantly higher values than the HTEM. Furthermore, on examining the the RMSEs for the noise variables (Figure \ref{fig:rmse_zero_htem_others}), we observe that the HTEM is able to identify the zero coefficients very well,  with the corresponding RMSEs close to 0. For the signals (Figure \ref{fig:rmse_nonzero_htem_others}), the accuracy of estimation of the HTEM is at least comparable with the rest of the methods. To summarize, in a sparse framework, the HTEM seems to be particularly effective in identifying the noise variables, and often has an overall superior performance compared to several competing methods.
\begin{table}
\centering
\caption{RMSEs of all the regression coefficients for the HTEM and the other studied competing methods under different scenarios, averaged over 100 replicates. The smallest value for each scenario is highlighted in bold font.}
\begin{tabular}{ccccc}
    \hline
    & HTEM & MONOT & MONON & HBL \\
    \hline
    Scenario (I) (Hyperbolic) & \textbf{0.064} & 0.134 & 0.144 & 0.228 \\
    Scenario (II) (Normal) &  \textbf{0.032} & 0.046 & 0.047 & 0.137 \\
    Scenario (III) (Student-t) &  \textbf{0.040} & 0.048 & 0.122 & 0.153 \\
    Scenario (IV) (Hyperbolic) &  \textbf{0.074} & 0.157 & 0.166 & 0.176 \\
    Scenario (V) (Normal) &  \textbf{0.038} & 0.061 & 0.061 & 0.109 \\
    Scenario (VI) (Student-t) &  \textbf{0.043} & 0.062 & 0.138 & 0.116 \\
   \hline
\end{tabular}
\label{tab:rmse_htem_others}
\end{table}

We also assess the efficacy of the HTEM in variable selection by comparing the same with the competing methods involving spike and slab priors, i.e., the MONOT and the MONON. For every $j = 1, 2, ..., p$, we look at the inclusion indicator $\gamma_j$, which is estimated using the marginal posterior inclusion probability $P(\gamma_j=1 | \bm{Y})$ as
\begin{equation}
    \hat{\gamma}_j = 1 \iff P(\gamma_j=1 | \bm{Y}) \geq \lambda,
\end{equation}
where $\lambda \in (0, 1)$ is a suitable threshold. Since the marginal posterior inclusion probabilities are analytically intractable, we use the corresponding MCMC sample proportions as their estimates. The next question relates to the choice of the threshold $\lambda$. The guiding instrument in our case is the marginal Bayes factor (\cite{Kass:Raft:1995, Ghos:Ghat:2015}) for $\gamma_j = 1$ versus $\gamma_j = 0$, defined as
\begin{equation}
    \mathrm{BF}_j = \dfrac{P(\gamma_j=1 | \bm{Y}) / P(\gamma_j=0 | \bm{Y})}{P(\gamma_j=1) / P(\gamma_j=0)},
\end{equation}
which is the ratio of the posterior odds to the prior odds of $\gamma_j = 1$. Aligning with the suggestions by \cite{Kass:Raft:1995}, a marginal Bayes factor exceeding 3.2 would indicate substantial evidence in favor of including the $j$th covariate in the model. In this context, a possible option is to employ the median probability model (MPM) with $\lambda = 0.5$, as proposed by \cite{Barb:Berg:2004}. This corresponds to a marginal Bayes factor threshold of 1 under a uniform prior on the inclusion probability $\tilde{\pi}$. However, such an approach appears to provide a bit stringent condition of variable selection for the sparse framework with a $\mathrm{Beta}(1, \sqrt{p})$ prior on $\tilde{\pi}$, considered in our study, because using the MPM with our choice of a $\mathrm{Beta}(1, \sqrt{p})$ prior leads to a very high threshold of 10 for the marginal Bayes factor. Therefore, while retaining the cutoff for marginal Bayes factor at 3.2, we use the prior under our sparse framework, leading to a threshold of $\lambda = 0.24$ for the marginal posterior inclusion probabilities. Using the subsequent estimates $\hat{\gamma}_j$s, we compute the true positive rates (TPRs) and true negative rates (TNRs) against the true inclusion indicators, in order to assess the variable selection performances of the methods. It is to be noted that TPR denotes the proportion of signals correctly captured in a model, among all the true signals. Similarly, TNR refers to the proportion of correctly dropped noise variables among all the true noise variables. As such, both TPR and TNR should be equal to 1 in an ideal case of perfect detection of signals and noise variables. For our study, the TPRs and TNRs obtained by the HTEM, the MONOT and the MONON are summarized in Table \ref{tab:tprtnr_htem_others}, based on the 100 simulated replications. The TPR values in Table \ref{tab:tprtnr_htem_others} are very close to 1 and are nearly comparable for all the methods under each of the six scenarios. On the other hand, the HTEM shows a significant gain in TNR. On a closer look, under scenarios corresponding to the mixed signals (Scenarios (IV), (V) and (VI)), the gain in TNR by the HTEM appears to be even more notable. In other words, the proposed HTEM shines in identifying the noise variables and appears to be slightly less effective than the other methods in identifying the signals, thereby corroborating the observations based on RMSEs of estimated regression coefficients.
\begin{table}
\centering
\caption{TPRs (TNRs) of all the regression coefficients for the HTEM and the other studied competing methods under different scenarios (using a threshold $\lambda$ = 0.24 for marginal inclusion probabilities). The values are averaged over 100 replications.}
\begin{tabular}{ccccc}
    \hline
    & HTEM & MONOT & MONON \\
    \hline
    Scenario (I) (Hyperbolic) &  1.000 (0.995) & 1.000 (0.680) & 1.000 (0.665) \\
    Scenario (II) (Normal) &  1.000 (0.998) & 1.000 (0.917) & 1.000 (0.916) \\
    Scenario (III) (Student-t) &  1.000 (0.990) & 1.000 (0.933) & 0.994 (0.715) \\
    Scenario (IV) (Hyperbolic) &  0.818 (0.993) & 0.985 (0.208) & 0.980 (0.237) \\
    Scenario (V) (Normal) &  0.953 (0.996) & 1.000 (0.771) & 1.000 (0.771) \\
    Scenario (VI) (Student-t) &  0.958 (0.988) & 1.000 (0.756) & 0.995 (0.437) \\
    \hline
\end{tabular}
\label{tab:tprtnr_htem_others}
\end{table}

\begin{figure}[!ht]
\centering
    \includegraphics[scale=0.75]{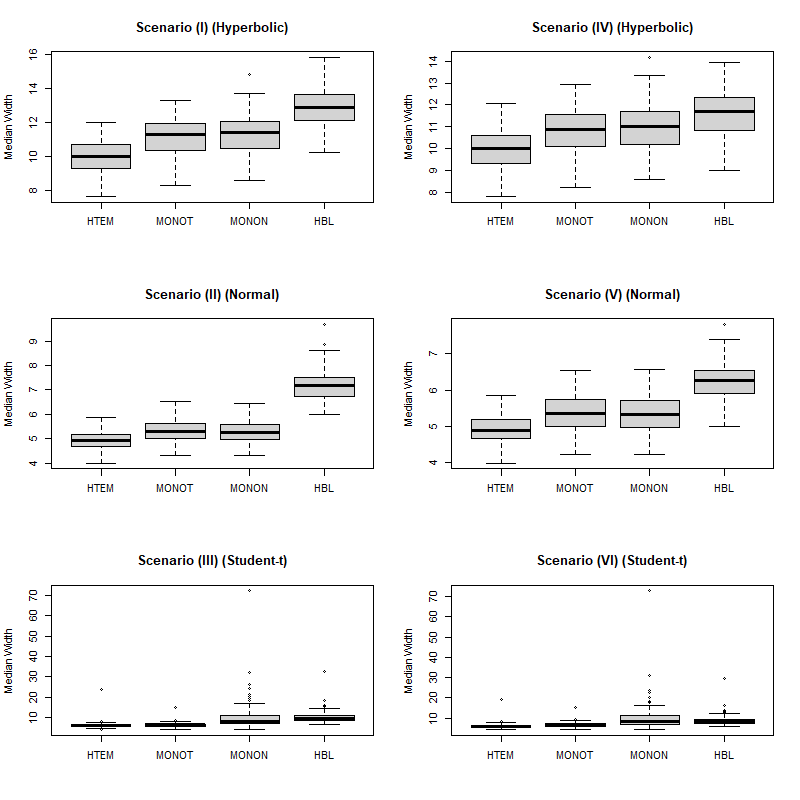}
\caption{Median widths of prediction intervals for the HTEM and the other studied competing methods under different scenarios. The boxplots are based on 100 replicates.}
\label{fig:pimw_htem_others}
\end{figure}

We next study the predictive performance of the methods. We construct 90\% prediction intervals for 1,000 out-of-sample points as in Section \ref{sec:simul_hem_htem_tem}. The empirical coverages are compiled in Table \ref{tab:picp_htem_others}, while boxplots of median widths are reproduced in Figure \ref{fig:pimw_htem_others}. The HTEM creates prediction intervals with a consistent coverage of about 90\% across all the scenarios, as evident from Table \ref{tab:picp_htem_others}. The MONON and the HBL produce an overcoverage under Student-$t$ errors, while the MONOT performs quite similar to the HTEM. Moreover, from Figure \ref{fig:pimw_htem_others}, it is revealed that the steady coverage of the HTEM is achieved with shorter intervals than the other methods. Even under true Student-$t$ errors, the results for the HTEM are almost similar to those of the MONOT. Accordingly, a prospective advantage in using the HTEM is apparent in terms of predictions as well.
\begin{table}
\centering
\caption{Empirical coverage probabilities of 90\% prediction intervals for the HTEM and the other studied competing methods under different scenarios, averaged over 100 replicates.}
\begin{tabular}{ccccc}
    \hline
    & HTEM & MONOT & MONON & HBL \\
    \hline
    Scenario (I) (Hyperbolic) & 0.897 & 0.891 & 0.890 & 0.886 \\
    Scenario (II) (Normal) & 0.899 & 0.906 & 0.904 & 0.899 \\
    Scenario (III) (Student-t) & 0.898 & 0.900 & 0.922 & 0.922 \\
    Scenario (IV) (Hyperbolic) & 0.895 & 0.874 & 0.873 & 0.888 \\
    Scenario (V) (Normal) & 0.894 & 0.893 & 0.891 & 0.888 \\
    Scenario (VI) (Student-t) & 0.891 & 0.893 & 0.914 & 0.911 \\
   \hline
\end{tabular}
\label{tab:picp_htem_others}
\end{table}

\subsection{Detection of True Error Density}

A pertinent query is exploring the ability of our proposed method to detect the true error model. We recall that in the proposed HTEM, an indicator $\alpha$ is introduced to switch between hyperbolic and Student-$t$ densities. In particular, the hyperbolic density is chosen for $\alpha = 0$, while $\alpha = 1$ denotes the Student-$t$ density. Moreover, the normal distribution can be captured under both $\alpha = 0$ and $\alpha = 1$, for large values of $\eta$. Accordingly, it is reasonable to look at the posterior probability $P(\alpha=0|\bm{Y})$ of a hyperbolic error model. Considering the behavior of the densities with respect to $\alpha$, we would typically expect $P(\alpha=0|\bm{Y})$ to be close to 1 and 0 respectively, depending on whether the true errors are generated from hyperbolic or Student-$t$ distributions. For true normal errors, since both $\alpha=0$ and $\alpha=1$ can lead to a normal density under large $\eta$, $P(\alpha=0|\bm{Y})$ is likely to assume a value of about 0.5.

\begin{figure}[!ht]
\centering
    \includegraphics[scale=0.75]{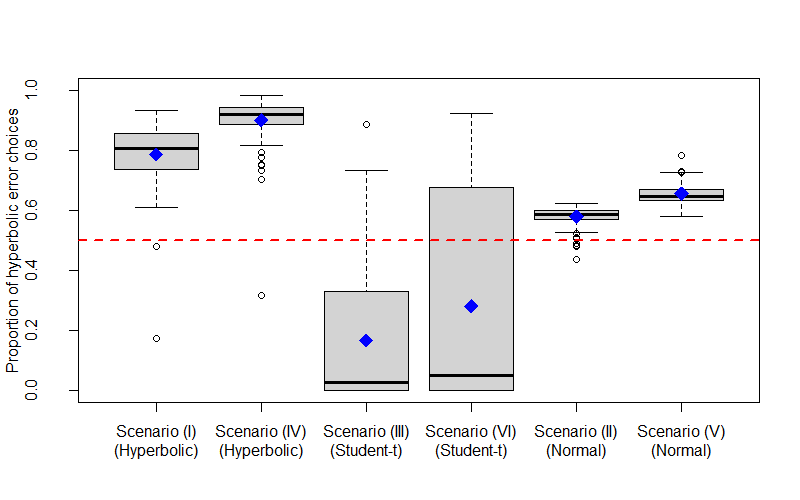}
\caption{Estimated posterior probability of choosing the hyperbolic error model by the HTEM under different scenarios. The diamonds indicate the average proportions and the dashed line marks a proportion of 0.5, corresponding to the prior probability. The boxplots and averages are based on 100 replicates.}
\label{fig:htem_hypsel}
\end{figure}

In the present article, we study the proportion of times a hyperbolic density has been chosen by the HTEM, which is the Monte Carlo estimate of $P(\alpha=0|\bm{Y})$. For each of the scenarios, the corresponding MCMC sample proportion is used as an estimate. The results are displayed in Figure \ref{fig:htem_hypsel}. Under Scenarios (I) and (IV), when the true errors are generated from a hyperbolic distribution, the posterior probability of a hyperbolic model, $P(\alpha=0|\bm{Y})$, is estimated to be much higher than the corresponding prior probability 0.5. Similarly, under Scenarios (III) and (IV), which correspond to true Student-$t$ errors, the posterior probability of the Student-$t$ error model, given by $1-P(\alpha=0|\bm{Y})$, is usually estimated to be higher than 0.5, the prior probability. However, the (estimated) posterior probability of the correct error model, is higher under true hyperbolic errors in Scenarios (I) and (IV), than those under Scenarios (III) and (VI). As far as the true normal errors in Scenarios (II) and (V) are concerned, the estimate of the posterior probability of a hyperbolic model is close to or slightly above 0.5. In other words, all the scenarios yield sensible results, which further empirically justifies the usefulness of the proposed HTEM, as discussed in Section \ref{sec:htem_posterior}.

\section{Real Data Analysis} \label{sec:data}

In this section we demonstrate the performance of the HTEM when applied to some real life datasets. In particular, we explore two datasets, namely, the well known Boston Housing dataset, and a recent dataset on salaries of NBA players.

We divide each dataset into a training and a test set, following the split ratios mentioned explicitly in Sections \ref{sec:data_boston} and \ref{sec:data_nba}. The process is replicated 100 times by creating 100 different random splits of the datasets, while maintaining the same split ratio across the replications for each dataset. This is done in order to curtail the sensitivity of the results to a specific choice of split.

Our investigation examines two aspects. The first point of interest is to evaluate the posterior behaviors of $\alpha$ and $\eta$ in the HTEM. The other goal is to compare the predictive efficiency of the HTEM with that of the MONOT, the MONON and the HBL. To that end, we implement the Gibbs samplers for all the techniques on the training sets for 100,000 iterations post burn in (10,000), with the same hyperparameters as in the simulation study in Section \ref{sec:simul}. The error distribution indicator $\alpha$ is examined through the proportion of times a hyperbolic density gets chosen among the MCMC samples. On the other hand, the MCMC sample mode of $\eta$ is studied to get an idea about the tail heaviness. After fitting the models to a training set, the predictive efficiencies are assessed using median absolute deviation (MeAD), which is defined as the median of absolute differences between the predicted and the actual values of the corresponding test samples. We also examine 90\% prediction intervals through empirical coverage probabilities and median widths.

\subsection{Boston Housing Dataset} \label{sec:data_boston}

The Boston Housing dataset from the \texttt{MASS} package in R, consisting of 506 observations on 13 neighborhood characteristics as regressors and median house value as response variable, often serves as a benchmark example with heavy-tailed errors. As a preprocessing step, we undertake a logarithmic transformation on the response variable to ensure an approximate symmetry among the residuals. The dataset contains 13 variables, all of which appear to be significant based on the least squares model. We make the problem more interesting for a variable selection framework, by incorporating many noise variables in the design matrix, as in \cite{Ghos:Reit:2012}. In particular, we generate 100 noise variables independently from a standard normal distribution, resulting in a total of $p = 113$ regressors. We compare the different methods to test how effectively they can drop the noise variables from the model, which in turn could lead to improved predictive performance.

\begin{figure}[!ht]
\centering
    \includegraphics[scale = 0.75]{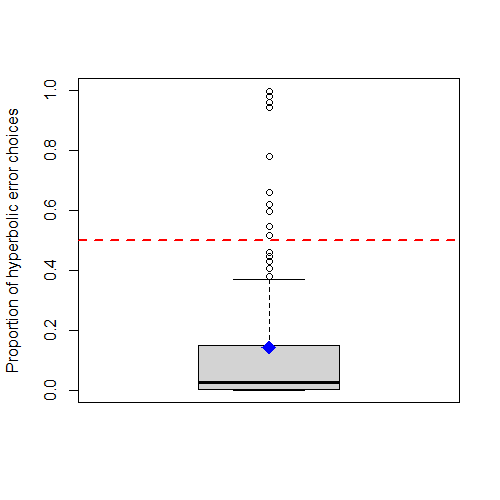}
    \caption{Estimated posterior probability of choosing the hyperbolic error model by the HTEM, for the Boston Housing dataset. The diamond indicates the average proportion and the dashed line marks a proportion of 0.5, corresponding to the prior. The boxplot and average are based on 100 replicates.}
\label{fig:boston_errpost}
\end{figure}

\begin{figure}[!ht]
\centering
    \includegraphics[scale=0.75]{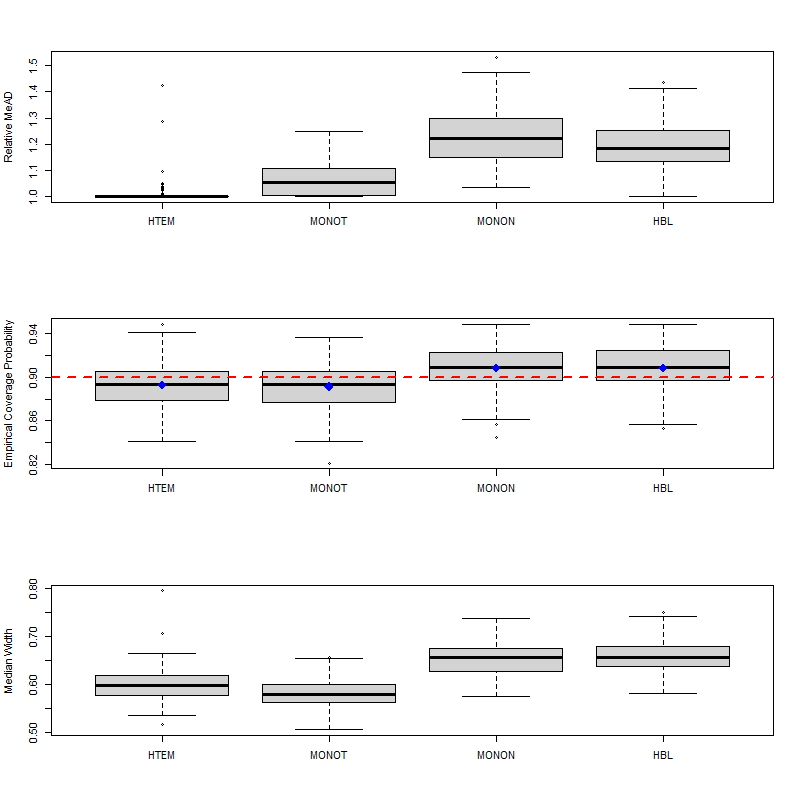}
\caption{Relative MeADs (top panel), empirical coverage probabilities (middle panel) and median widths (bottom panel) of 90\% prediction intervals for the HTEM and the other studied competing methods for the Boston Housing dataset. In the top panel, relative MeAD of any method is its MeAD relative to that of the method having the smallest MeAD. In the middle panel, the diamonds indicate the mean empirical coverages and the dashed line marks a coverage of 90\%. The boxplots and means are based on 100 replicates.}
\label{fig:boston_pred}
\end{figure}

A 50\%-50\% split ratio is used to divide the dataset into training and test samples for each of the 100 replications. Figure \ref{fig:boston_errpost} conveys an overall inclination towards choosing Student-$t$ errors. We find that the MCMC sample mode of $\eta$ lies at a value of 2 very consistently across the replications, except for a handful of fluctuations to even smaller values.
The boxplots of the metrics pertaining to predictive efficiency are presented in Figure \ref{fig:boston_pred}, based on the 100 replicates.
Based on the first panel in Figure \ref{fig:boston_pred}, the HTEM has the smallest MeAD. As a heavy-tailed technique with spike and slab priors, the MONOT acts as the closest competitor to the HTEM with slightly larger MeAD values. The HBL produces larger values of MeAD, followed by the non-robust MONON. In terms of prediction intervals, all the four methods perform quite similarly in maintaining the empirical coverage, with the MONON and the HBL yielding slightly more conservative intervals, as evident from the second panel in Figure \ref{fig:boston_pred}. The third panel in that plot shows that the HTEM has median widths comparable to the MONOT, which typically prompts the shortest intervals. The intervals corresponding to the MONON and the HBL are much wider. Considering the higher preference for Student-$t$ errors in Figure \ref{fig:boston_errpost}, for the Boston housing dataset, the predictive performance of different methods for this dataset seems to concur with the findings in the simulation study.

\subsection{NBA Player Salaries Dataset} \label{sec:data_nba}

We now turn to the NBA player salaries dataset. The dataset contains the salaries of 467 basketball players participating in the 2022-23 season of NBA, along with 49 other features, including teams, player positions, and several player statistics and performance indicators. Since teams and positions are categorical variables, we create dummy variables for the respective categories. The categories with 5 or less observations are dropped to avoid potential complications that might arise due to inadequate inclusion of such rare categories in the training set. Furthermore, we exclude the observations with missing data. After performing these data processing steps, a total of 374 observations are left, with a response variable (salary) and $p = 81$ predictors. The resulting dataset exhibits multicollinearity among the covariates, thereby making the problem more intricate than the simulation study as well as the Boston Housing example in the preceding sections. In addition, to control the unstable residual variance caused due to the extremely large salary figures, the salaries are divided by a factor of $10^{6}$, and to maintain a rough symmetry among the residuals, the salaries are subjected to a square root transformation thereafter.

For each of the 100 replications, 196 randomly selected observations are included in the training sample to maintain a reasonable share of all the categories, while holding onto the remaining 178 observations in the test sample. We first look at the choice of the error distribution by the HTEM, along with the estimation of the corresponding tail heaviness parameter, based on the observed data.
\begin{figure}[!ht]
\centering
    \includegraphics[scale = 0.75]{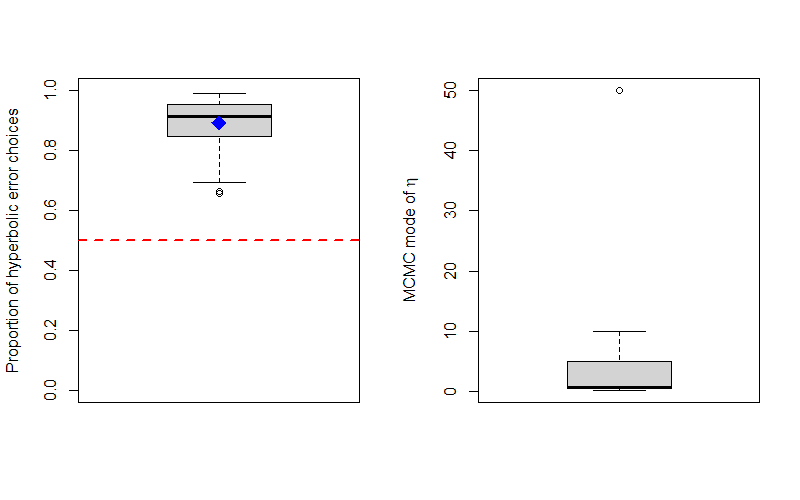}
    \caption{(Left panel) Estimated posterior probability of choosing the hyperbolic error model by the HTEM, for the NBA Player Salaries dataset. The diamond indicates the average proportion and the dashed line marks a proportion of 0.5, corresponding to the prior probability. (Right panel) MCMC sample modes of the tail heaviness parameter $\eta$ under the HTEM. Both the boxplots and the average are based on 100 replicates.}
\label{fig:nba_errpost}
\end{figure}
The boxplot in the left panel in Figure \ref{fig:nba_errpost} indicates remarkable priority to the hyperbolic error model. The boxplot in the right panel in Figure \ref{fig:nba_errpost} shows that the posterior modes of $\eta$ are mostly located around the smaller values. In other words, our proposed model prefers a hyperbolic error model, typically with heavy tails.
\begin{figure}[!ht]
\centering
    \includegraphics[scale=0.75]{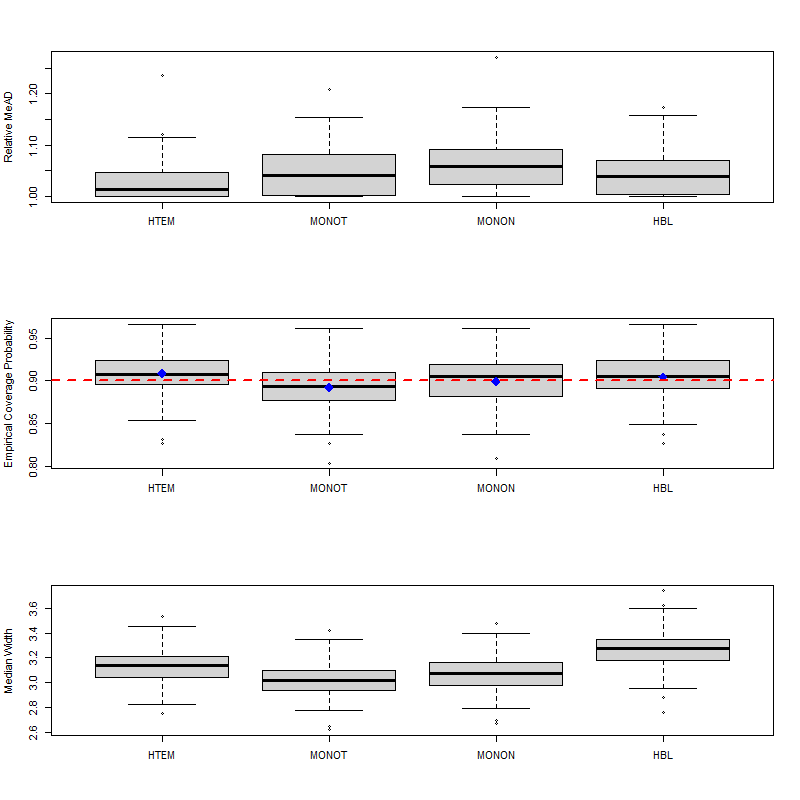}
\caption{Relative MeADs (top panel), empirical coverage probabilities (middle panel) and median widths (bottom panel) of 90\% prediction intervals for the HTEM and the other studied competing methods for the NBA Player Salaries dataset. In the top panel, relative MeAD of any method is its MeAD relative to that of the method having the smallest MeAD. In the middle panel, the diamonds indicate the mean empirical coverages and the dashed line marks a coverage of 90\%. The boxplots and means are based on 100 replicates.}
\label{fig:nba_pred}
\end{figure}
From Figure \ref{fig:nba_pred}, all the methods seem to maintain a nearly 90\% coverage, with the MONOT and the MONON having slightly shorter prediction intervals than the other two methods. Similar to the Boston Housing dataset, the HTEM continues to have slightly lower MeAD values compared to the remaining competing models.

\section{Conclusion} \label{sec:conc}

The primary objective of the article has been to design a fully Bayesian model averaging approach to tackle variable selection under heavy-tailed errors. Transcending the thin-tailed normal error models, we have explored a more general class of error densities, viz., the hyperbolic density, in the modeling framework. The alluring nature of this distribution is its adaptability to varying extent of tail heaviness. Additionally, we have also accounted for the potentially thicker-tailed Student-$t$ distribution. Using a mixture of these two densities, we have created an amalgamated error structure. Together with that, we have devised a new model (HTEM) that incorporates the gold standard prior for Bayesian variable selection, specifically, the spike and slab priors for regression coefficients. It has been demonstrated, both illustratively and intuitively, that the HTEM is a more flexible approach than working with its components, the HEM or the TEM, alone. Promising results have been obtained for the HTEM when compared to other state-of-the-art methods through several simulation studies and two real datasets. Precisely, the effectiveness of our proposed technique is appealing in terms of variable selection, regression coefficient estimation, and prediction. The HTEM has also demonstrated competitive performance with several well-known Bayesian variable selection methods in uncertainty quantification, particularly in generating prediction intervals with good frequentist coverage.

We conclude this article with some promising and important directions for  future work. In this article, we have considered two well-known symmetric members of the family of generalized hyperbolic distributions (\cite{Gnei:1997}), namely, the hyperbolic and the Student-$t$ distributions. Potentially, the parameter $p$ in the $\mathrm{GIG}(p, a, b)$ distribution, the mixing density for scale mixture of normals, can be given a flexible prior to accommodate a richer family of heavy-tailed error distributions. Another direction is to consider asymmetric error distributions by putting a prior on $\mu$ in (\ref{eq:hyperbdensityorig}). Some authors (\cite{Gagn:Desg:Beda:2020, Hamu:Irie:Suga:2022}) have theoretically established posterior robustness. Empirical assessment of the robustness of the HTEM in terms of well known yardsticks of robustness, such as the influence function (\cite{Kawa:Hash:2023}), may be an interesting question to address in the future. Other viable directions include scenarios, where the number of predictors and/or the sample size is much larger than the ones considered here. In a very high-dimensional framework, standard MCMC algorithms can become trapped in local modes. There are many possibilities that include 1) exploring alternative MCMC algorithms that are customized for highly multimodal distributions, 2) simplifying the problem by modifying the goal to estimate the posterior mode(s), instead of the entire posterior distribution, or 3) considering alternative continuous shrinkage priors for variable selection such as the horseshoe prior, to name a few. While a very large sample size is desirable for estimation, it can reduce the speed of algorithms substantially, which can create a different kind of computational challenge. In such scenarios, alternative algorithms based on divide-and-conquer methods or those based on subsampling could prove to be useful. Last but not the least, exploring the stability of the algorithms and results under varying hyperparameter choices is also worth considering.

\section*{Acknowledgments}
Joyee Ghosh's research was supported by NSF Grant DMS-1612763. Any opinions, findings, and conclusions or recommendations expressed in this publication are those of the author and do not necessarily reflect the views of the National Science Foundation. The authors thank Dr.~David Dunson for helpful discussions about robust regression.

\bibliographystyle{apalike}
\bibliography{jcgs-htem}

\section*{Appendix: Some Useful Distributions} \label{app:densities}

We briefly review a few distributions that are frequently encountered in this article, mainly for clarity on the parametrizations of their density functions used herein.

\begin{enumerate}

\item
Let $\mathrm{GIG}(p, a, b)$ denote the generalized inverse Gaussian (GIG) distribution with the density
\begin{equation}
    f_{\mathrm{GIG}}(x | p, a, b) = \dfrac{(a/b)^{p/2}}{2 K_p(\sqrt{ab})} x^{p-1} e^{-(ax + b/x)/2},
    \quad x > 0,
\label{eq:gigdensity}
\end{equation}
where $a, b > 0$, $p \in \mathbb{R}$ and $K_p$ is a modified Bessel function of the second kind.

\item
The gamma distribution $\mathrm{Gamma}(a, b)$ has a density of the form
\begin{equation}
    f_{\mathrm{Gamma}}(x | a, b) = \dfrac{b^a}{\Gamma(a)} x^{a-1} e^{-bx},
    \quad x > 0,
\label{eq:gammadensity}
\end{equation}
where $a > 0$ and $b > 0$ respectively indicate the shape and the rate parameters.

\item
Likewise, the inverse gamma distribution $\mathrm{IGamma}(a, b)$ has the density
\begin{equation}
    f_{\mathrm{IGamma}}(x | a, b) = \dfrac{b^a}{\Gamma(a)} x^{-a-1} e^{-b/x},
    \quad x > 0,
\label{eq:invgammadensity}
\end{equation}
where $a > 0$ and $b > 0$ respectively indicate the shape and the scale parameters.

\item
The probability mass function of the discrete uniform distribution $\mathrm{DiscUniform}(\mathcal{G})$ on a finite set of real values $\mathcal{G}$ is expressible as
\begin{equation}
    f_{\mathrm{DiscUniform}}(x) = \dfrac{1}{|\mathcal{G}|},
    \quad x \in \mathcal{G},
\label{eq:discuniformdensity}
\end{equation}
where $|\mathcal{G}|$ represents the number of elements in $\mathcal{G}$.

\end{enumerate}

\end{document}